\documentclass[]{article}
\usepackage{graphics}
\usepackage{endnotes}

\newcommand{\bc}{\begin{center}}
\newcommand{\ec}{\end{center}}
\newcommand{\beq}{\begin{equation}}
\newcommand{\eeq}{\end{equation}}
\newcommand{\bea}{\begin{eqnarray}}
\newcommand{\eea}{\end{eqnarray}}
\newcommand{\ba}{\begin{array}}
\newcommand{\ea}{\end{array}}

\def\eg{\emph{e.g., }}
\def\ie{\emph{i.e., }}
\def\etal{\emph{et al.}}

\newcommand{\sgn}[1]{\ensuremath{\mathrm{sgn}(#1)}}

\def\eps{\epsilon}

\def\J{{\bf J}}
\def\F{{\bf F}}

\def\ch{\hat{c}}
\def\phih{\hat{\phi}}
\def\psih{\hat{\psi}}

\def\Fh{\hat{\bf F}}

\def\cb{\bar{c}}

\def\Fb{\bar{\bf F}}

\begin{document}
\bibliographystyle{achemso}

\title{Surface Conservation Laws at Microscopically Diffuse Interfaces} 

\author{ 
  Kevin T. Chu\footnotemark[1] $^,$\footnotemark[2] $^,$\footnotemark[3]
  and Martin Z. Bazant\footnotemark[3] \\ \ \\
\it Department of Mechanical and Aerospace Engineering,
Princeton University, Princeton, NJ 08544  \\
\it Department of Mathematics, Massachusetts Institute of Technology,
Cambridge, MA 02139 
}

\date{ February 5, 2007 }  

\maketitle

\renewcommand{\thefootnote}{\fnsymbol{footnote}}
\footnotetext[1]{Corresponding author.  E-mail: ktchu@princeton.edu}
\footnotetext[2]{Princeton University}
\footnotetext[3]{Massachusetts Institute of Technology}
\renewcommand{\thefootnote}{\arabic{footnote}}

\begin{abstract}
In studies of interfaces with dynamic chemical composition, bulk and 
interfacial quantities are often coupled via surface conservation laws of 
\emph{excess} surface quantities.  While this approach is easily justified 
for microscopically sharp interfaces, its applicability in the context of 
microscopically \emph{diffuse} interfaces is less theoretically 
well-established.  Furthermore, surface conservation laws (and interfacial
models in general) are often derived phenomenologically rather than 
systematically.
In this article, we first provide a mathematically rigorous justification for 
surface conservation laws at diffuse interfaces based on an asymptotic 
analysis of transport processes in the boundary layer and derive general 
formulae for the surface and normal fluxes that appear in surface 
conservation laws.  Next, we use non-equilibrium thermodynamics
to formulate surface conservation laws in terms of chemical potentials and 
provide a method for systematically deriving the structure of the 
interfacial layer.  Finally, we derive surface conservation laws for 
a few examples from diffusive and electrochemical transport. 

\end{abstract}

\section{Introduction}
It is well-known that the chemical composition of an interface plays
an important role in its dynamics (both its motion and the evolution of 
its chemical composition).  A few interesting and important examples 
include the effect of surfactants on surface 
tension~\cite{ferri2000, mulqueen2001b,mulqueen2001a}, 
the charging dynamics of electrochemical double 
layers~\cite{dukhin1969,shilov1970,hinch1983,hinch1984, bazant2004}, 
electrokinetic phenomena~\cite{bikerman1940, deryagin1969,saville1997,
ramos1998, ramos1999, ajdari2000, green2000,gonzalez2000,rubinstein2000,
rubinstein2001,iceo2004a,iceo2004b} and 
soap film dynamics~\cite{chomaz2001,couder1989}.  Unfortunately,
theoretical modeling of the interaction between chemical composition and 
interfacial dynamics can be quite challenging because it involves coupling
dynamics at the macroscopic scale away from the interface with a microscopic
model for the interfacial region.  Sometimes it may not even be
clear which microscopic model is appropriate for a particular problem.  
Many successful approaches based on the notion of an excess surface 
concentration~\cite{hunter_book, lyklema_book_vol_1} have been developed 
and applied to a wide-range of problems.  
While this approach is well-founded for microscopically sharp interfaces, 
the theoretical foundations for this methodology in the context of 
microscopically diffuse interfaces (especially for dynamic interfaces) 
do not appear to have been thoroughly explored.  

For microscopically diffuse interfaces (\ie interfaces where 
the chemical composition smoothly varies across the interface), the 
notion of a surface excess concentration and dynamics of surface excess 
quantities can be rigorously justified via asymptotic analysis.  In this 
article, we present a general procedure for deriving surface conservation 
laws (including derivation of the structure of the interfacial layer) that 
couple dynamics in the bulk with dynamics of the interfacial region
and demonstrate the application of our formulation to example problems
in diffusive and electrochemical transport.
Our results are only valid in ``sharp interface'' limit (\ie when the 
distance over which the volume density of the diffuse species varies is 
small relative to macroscopic length scales of the problem).  
Fortunately, this limit is the only one that is physically meaningful -- if 
an interface is too diffuse, it may not be appropriate to treat it as an 
interface in the first place.

Previous work involving surface conservation laws focuses primarily on deriving 
the appropriate surface conservation law in the context of specific 
problems~\cite{deryagin1969, dukhin1969, shilov1970, hinch1983,hinch1984}.
In contrast, our derivation is much more general and applicable to a wide
range of transport problems.  We only require that, near the surface, the 
flux for the transport process scales in the same manner as a linear 
combination of gradients of field variables.  
Fortunately, these types of fluxes are very common in transport 
theory~\cite{degroot_book}.  
We emphasize that the formulation of surface conservation laws we present 
is \emph{not} based on ad-hoc physical arguments; rather, they are a 
direct consequence of the asymptotic analysis.  
Moreover, the generality of our formulation allows us to derive surface 
conservation laws for nonlinear transport processes when the bulk is far 
from equilibrium (\eg electrochemical transport of around metallic particles at
large applied fields) and for systems with complex interfacial structure.
To the authors' knowledge, surface conservation laws for these types of 
problems have not been previously derived.

\subsection{Dimensionless Formulation of Equations}
To facilitate the mathematical analysis presented in this article, it is 
convenient to examine all equations in nondimensional form.  To avoid 
confusion when reintroducing dimensions, let us take a moment to 
fix the physical scales used in the nondimensionalization process.  
At the macroscopic scale, the spatial coordinates, $(X,Y,Z)$, are scaled by 
the the characteristic size of the system, $L$.  At the microscopic scale, 
the tangential, $(x,y)$, and normal, $z$, spatial coordinates are scaled 
by $L$ and $\delta$, respectively, where $\delta$ is the characteristic 
thickness of the interfacial layer.  Time, $t$, is scaled by the bulk
diffusion time scale, $\tau = L^2/D$.
Concentrations, $c_i$, are nondimensionalized by taking the reference 
concentration to be the maximum physically realizable concentration, 
$c_{max}$.  Bulk, $\F_i$, and surface fluxes, $\J_i$, are scaled using 
$c_{max}L/\tau$ and $c_{max}L^2/\tau$, respectively.  The electric 
potential, $\phi$, is scaled by the thermal voltage $k_B T/e$, where
$k_B$ is Boltzmann's constant and $e$ is the absolute charge of an electron.  
Finally, all thermodynamic energy variables (\eg $U$ and $F$), and entropy, 
$S$, are scaled using $k_B T$ (the thermal energy) and $k_B$, respectively.  
The dimensional form of any of the equations in the remainder of this article 
can be obtained by first replacing each dimensionless variable with its
associated dimensional variable divided by the appropriate characteristic 
scale and then multiplying the entire equation by an appropriate physical 
scale to restore units to the equation\endnote{Note that differential 
operators will also need their units restored.}.  For instance, if the 
dimensionless concentration, $c$, appears in a dimensionless equation, it 
should be replaced by $C/c_{max}$ when re-dimensionalizing the equation.
\begin{table}
\caption{\label{tab:summary_of_scales}  Summary of physical scales
used to nondimensionalize equations.} 

\begin{center}
\begin{tabular}{ccc}
Physical Quantity & Variable(s) & Physical Scale \\
\hline \\
spatial coordinate & $X,Y,Z,x,y$ & $L$ (macroscopic length scale) \\
spatial coordinate & $z$ & $\delta$ (microscopic length scale) \\
time & $t$ & $\tau=L^2/D$ (bulk diffusion time scale) \\
concentration & $c_i$ & $c_{max}$ (maximum physically realizable 
  concentration)\endnote{$c_{max} = 1/a^3$, where $a^3$ is the characteristic 
volume of each solute/solvent particle.} \\
bulk flux & $\F_i,J^i_n$ & $c_{max}L/\tau$ \\
surface flux & $\J^i_s$ & $c_{max}L^2/\tau$ \\
electric potential & $\phi$ & $k_B T/e$ (thermal voltage) \\
energy & $U,F$ & $k_B T$ (thermal energy) \\
entropy & $S$ & $k_B$ \\
\\ \hline \\

\end{tabular}
\end{center}
\end{table}

\section{Excess Surface Concentrations \label{sec:excess_surf_conc}}
When studying chemically active interfaces, the state of the interface 
is commonly described at macroscopic scales by specifying the 
\emph{excess} surface concentrations of all chemical species. 
Intuitively, excess surface concentrations are defined as the amount of
material per unit area of surface after the material in the ``bulk'' 
has been removed from the interfacial region.  However, this definition is 
physically and mathematically ambiguous -- in the region near the interface, 
how is one to distinguish between material that is part of the bulk and 
material that is part of the interface?  The ambiguity of this definition was 
recognized long ago by Gibbs~\cite{gibbs_1876,gibbs_1878,gibbs_works} 
and is typically dealt with by arbitrarily selecting a concentration, 
$C^*$, at some point within the interface (when viewed at a microscopic 
length scale) to be the reference ``bulk'' 
concentration~\cite{hunter_book, lyklema_book_vol_1,chattoraj_book}.  
Any deviation of the concentration near the interface from $C^*$ is treated 
as a contribution to the excess surface concentration.  In terms of this
reference concentration, the excess surface concentration is defined as the 
integral of $\left( C - C^* \right)$ over the thickness of the interface.

\begin{figure}[htb]
\bc
\scalebox{0.5}{\includegraphics{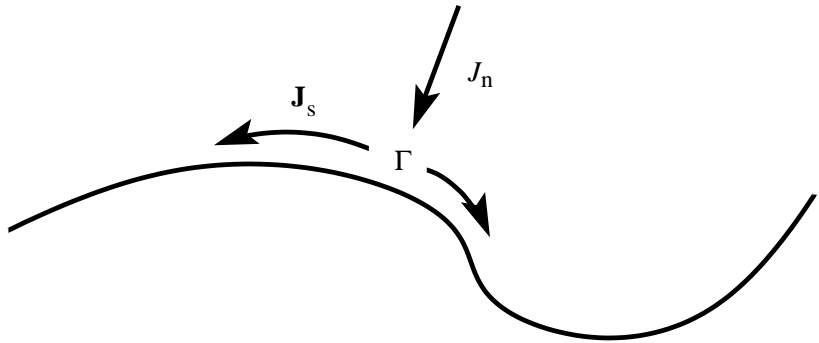}}
\begin{minipage}[h]{5in}
\caption[Schematic diagram of fluxes at a sharp interface]{
\label{fig:sharp_interface}
Schematic diagram of fluxes at an interface.  Note that the 
surface excess concentration $\Gamma$ and the surface fluxes $J_s$ 
are defined \emph{only} on the interface itself.
}
\end{minipage}
\ec
\end{figure}

While somewhat inelegant from a theoretical perspective, this formulation 
of excess surface concentration has been used quite successfully for studying 
interfaces of equilibrium systems.  For systems in thermal equilibrium, 
the excess surface concentrations can be related to other thermodynamic 
variables, such as bulk concentrations and other excess surface 
concentrations via adsorption isotherms~\cite{bard_book,ferri2000,hunter_book,
lyklema_book_vol_1,newman_book}.
However, when the dynamics of surface species is important (\eg fast 
adsorption-desorption kinetics, non-negligible surface-diffusion), 
isotherm models need to be replaced by surface conservation laws for the 
excess surface concentrations which provide dynamic coupling between 
bulk concentrations and excess surface concentrations: 
\bea
  \frac{\partial \Gamma}{\partial t} = 
   -\nabla_s \cdot \J_s + J_n.
  \label{eq:surface_conservation_law}
\eea
Here $\Gamma$, $\J_s$ and $J_n$ are the surface excess 
concentration, the surface flux and the normal flux, 
$\nabla_s$ denotes a surface derivative, and the sign on the normal flux 
is chosen to be positive when the flux is \emph{into} the boundary layer
(see Figure~\ref{fig:sharp_interface}).
In this situation, the thermodynamic definition of excess surface 
concentrations becomes unsatisfactory because all of the variables in
(\ref{eq:surface_conservation_law}) depend on implicitly on the choice
of $C^*$ whose governing equation and relationship to other macroscopic 
variables may be difficult to derive.

Fortunately, for many problems, there is a natural way to define $C^*$ in 
terms of bulk concentrations (which almost always have relatively 
straightforward governing equations).  Define $C^*$ to be the limit of
the corresponding bulk concentration, $C(x)$, as $x$ approaches the 
interface.  Unlike thermodynamic definitions of the excess concentration 
which are defined in terms of a reference concentration chosen at microscopic 
length scales and suffer from the difficulty of relating that reference 
concentration to macroscopic variables, the choice of $C^*$ just described 
is directly related to macroscopic variables because it is explicitly
defined in terms of those variables.  
Note that $C^*$ is not necessarily a constant; it may still be a function
of position along the surface of the interface.  

\section{Surface Conservation Laws}
Equation (\ref{eq:surface_conservation_law}) seems physically intuitive: the 
time rate of change in the surface concentration results from a combination 
of surface diffusion and flux from the bulk.  However, it is important to 
remember that the equation describes the evolution of \emph{excess} 
concentrations, not absolute concentrations.  Thus, $\J_s$ and $J_n$ 
must be carefully defined so that they contribute solely to changes in the 
excess concentration, not to changes in reference concentration $C^*$.

For microscopically sharp interfaces, such as the monolayer interfaces that 
arise in problems involving surfactants at liquid-gas 
interfaces~\cite{ferri2000,mulqueen2001b,mulqueen2001a} 
or the compact layer in electrochemical 
systems~\cite{bard_book,lyklema_book_vol_1}, $\J_s$ and $J_n$ are 
very simply defined because the bulk truly extends all the way to the 
interface.  As a result, the excess surface concentration is the amount
of material per unit surface area that resides precisely on the interface. 
In other words, the excess surface concentration is the \emph{absolute}
surface concentration.  Thus, the surface can truly be thought of as a 
distinct phase, and a conservation law argument based on a balance of 
fluxes into a ``control volume'' on the interface is mathematically valid.  
As a result, $J_n$ is the flux of bulk material normal to the surface
and $\J_s$ is the flux within the surface itself driven by transport
processes intrinsic to the interface (\eg surface diffusion in an lipid 
bilayer).

\begin{figure}[htb]
\bc
\scalebox{0.35}{\includegraphics{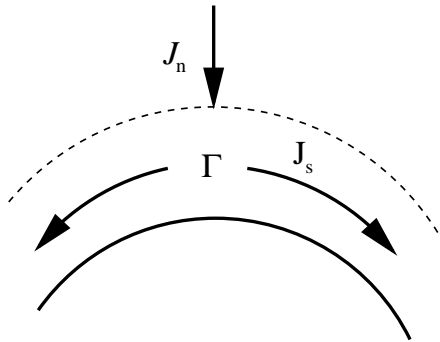}}
\begin{minipage}[h]{5in}
\caption[Schematic diagram of a microscopically diffuse interface]{
\label{fig:boundary_layer}
Schematic diagram of a microscopically diffuse interface.  
The dashed line represents the outer ``edge'' of the boundary 
layer, which is \emph{not} at a mathematically well-defined location.
Note that in addition to the fluxes within the boundary layer, it
is possible for the interface to possess fluxes that physically reside 
on the microscopic interface itself.  For some systems, these ``microscopic''
surface fluxes may provide a significant contribution to the total surface
flux in the boundary layer.}
\end{minipage}
\ec
\end{figure}

That surface conservation laws hold for microscopically sharp interfaces
is not surprising.  What is interesting is that $\J_s$ and $J_n$ can be 
defined in such a way that surface conservation laws also hold for 
microscopically \emph{diffuse} interfaces
(see Figure~\ref{fig:boundary_layer}). 
For instance, in their studies of the evolution of surface excess 
concentrations of electrolyte around colloid 
particles~\cite{deryagin1969,dukhin1969, shilov1970}, 
Dukhin, Deryagin, and Shilov made use of surface conservation laws for
excess ion concentration to couple ion transport in the bulk to the 
behavior of the electrical double layer.
A key observation about microscopically diffuse interfaces is
that surface transport within the boundary layer near the interface 
is driven by the \emph{same} transport processes that occur in the bulk.  
In contrast, at microscopically sharp interfaces, surface transport could 
potentially be completely different from bulk transport because the 
underlying physical processes driving surface transport (which are 
related to the specific nature of the interface) might be fundamentally
different.

Unfortunately, the derivation of surface conservation laws for 
microscopically sharp interfaces is not valid for diffuse interfaces
because the notion of a surface ``control volume'' is no longer well-defined. 
The main problem is the lack of a distinct separation between the bulk and 
the interfacial region.  Rather, there is a thin region near the interface 
over which concentrations vary rapidly.  Intuitively, what we would like 
to do is define surface quantities by directly integrating over the boundary 
layer.  This procedure, however, must be carried out carefully in order 
to obtain physically meaningful results.  The basic idea in deriving
surface conservation laws, which we shall elaborate on in the next section, 
is that all integrations over the boundary layer should involve only 
{\em excess} quantities; integration of absolute quantities leads to divergent 
results.

Before deriving the general form for $\J_s$ and $J_n$ involved in surface 
conservation laws for microscopically diffuse interfaces, we mention that 
(\ref{eq:surface_conservation_law}) neglects contributions
due to motion and deformation of the interface (\eg terms involving
convection in the normal direction and surface dilation~\cite{saville1997,
stone1990,wong1996}).  
Since our analysis sheds no new insight for these terms, we shall ignore 
them to simplify the discussion.

\subsection{Derivation of Equation (\ref{eq:surface_conservation_law}) for 
Microscopically Diffuse Interfaces}
To derive the form of the fluxes for surface conservation laws at
microscopically diffuse interfaces, our goal is to identify the appropriate 
definitions for $\J_s$ and $J_n$ (in terms of macroscopic variables) that 
will allow us to write surface conservation laws of the form 
(\ref{eq:surface_conservation_law}).  We begin by writing the conservation 
laws that govern the time evolution of a chemical species in the 
bulk:
\bea
  \frac{\partial c}{\partial t} = -\nabla \cdot \F,
  \label{eq:conservation_law_outer}
\eea
where $c$ is the concentration and $\F$ is the flux.  
For our derivation, we assume that there is no flux of material through the
interface\endnote{This assumption may be relaxed in situations where there 
are processes intrinsic to the interface (\eg chemical reactions).}
itself so that $\partial \F/\partial n = 0$ and that the spatial coordinates 
in (\ref{eq:conservation_law_outer}) have been nondimensionalized using 
a characteristic \emph{macroscopic} length scale.
In these units, the thickness of the boundary layer over which $c$ varies 
rapidly is $O(\eps)$, where $\eps \equiv \delta/L$ is the ratio of the 
interfacial and macroscopic length scales.  

The ``sharp interface'' limit is defined as the limit $\eps \rightarrow 0$ 
and is best studied via asymptotic analysis~\cite{bender_book,hinch_book} 
using (\ref{eq:conservation_law_outer}) for the {\em outer} equations and 
deriving the {\em inner} equations by appropriately rescaling the spatial 
coordinates.  In this discussion, we shall distinguish inner and outer 
variables by using the hat ($\hat{\ }$) and bar accents ($\bar{\ }$),
respectively.  Also, we shall use $(X,Y,Z)$ (upper case variables) and 
$(x,y,z)$ (lower case variables) to represent the outer and inner spatial 
coordinates, respectively.
Since the focus will be on the region immediately neighboring the 
surface, we shall take $(x,y)$ to be Cartesian coordinates tangential to 
the interface and $z$ to the coordinate normal to the interface.

Rescaling (\ref{eq:conservation_law_outer}) using the inner coordinates
$(x,y,\eps z) = (X,Y,Z)$, we find that the governing equations in the 
boundary layer are
\bea
  \frac{\partial \ch}{\partial t} = 
    -\nabla_s \cdot \Fh_s 
    - \frac{1}{\eps^2} \frac{\partial \Fh_n}{\partial z}
  \label{eq:conservation_law_inner},
\eea
where the subscripts $s$ and $n$ indicate tangential and normal 
components of the flux and divergence operator, respectively.
Note that in (\ref{eq:conservation_law_inner}), we have implicitly assumed 
that in changing to the inner coordinates, the normal derivative of the 
normal flux $\Fh_n$ picks up a factor of $1/\eps$ because
\beq
  \Fh_n (x,y,z) = \eps \Fb_n (X,Y,Z).  
  \label{eq:inner_outer_flux_relation}
\eeq
This relationship between the inner and outer normal fluxes is a direct 
consequence of our requirement that, near the interface, $\F$ scales in the 
same manner as a linear combination of gradients of macroscopic field 
variables:
\bea
  \F = \sum_i f_i\left(c_1, c_2, \ldots, c_n \right) \nabla c_i,
  \label{eq:flux_equation}
\eea
where $f_i$ are arbitrary functions of $c_1, c_2, \ldots, c_n$.
For fluxes of the form (\ref{eq:flux_equation}), the factor of $\eps$ 
in (\ref{eq:inner_outer_flux_relation}) arises from the fact that the 
flux itself involves derivatives of space.  We emphasize, however,
that for our general analysis, we only require that fluxes satisfy
(\ref{eq:inner_outer_flux_relation}) without regard for the origin of 
this relationship. 

In deriving (\ref{eq:surface_conservation_law}), the intuitive
idea of just integrating (\ref{eq:conservation_law_inner}) 
from $z = 0$ to $z = \infty$ is inadequate because it leads to 
divergent integrals which are physically and mathematically meaningless.
The key idea to keep in mind when asymptotically integrating over
boundary layers is that only \emph{excess} concentrations
are integrable, absolute concentrations are not.

Using this basic principle, we can systematically derive 
(\ref{eq:surface_conservation_law}) by starting with the excess 
(volume) concentration in the boundary layer:
\bea
  \gamma(x,y,z,t)  &=& \ch(x,y,z,t) - c^*(x,y,t),
\eea
where $c^*(x,y,t) \equiv \lim_{Z \rightarrow 0} \cb(X,Y,Z,t)$.
Notice that, as discussed in Section \ref{sec:excess_surf_conc}, we have 
chosen the reference concentration as limit of the bulk concentration as 
the interface is approached.
Taking the time derivative of this equation and using
inner and outer evolution equations, 
(\ref{eq:conservation_law_outer}) and (\ref{eq:conservation_law_inner}),
we find that
\bea
  \frac{\partial \gamma}{\partial t} &=&
  \left ( -\nabla_s \cdot \Fh_s 
          -\frac{1}{\eps^2} \frac{\partial \Fh_n}{\partial z} 
  \right )
  - \left (-\nabla_s \cdot \F^*_s 
           -\lim_{Z \rightarrow 0} \frac{\partial \Fb_n}{\partial Z} 
    \right),
  \label{eq:gamma_evolution_equation}
\eea
where $\F^* \equiv \lim_{Z \rightarrow 0} \Fb$.
To obtain an equation for the surface excess concentration $\Gamma$, 
we would like to integrate this equation over the entire boundary layer.
Unfortunately, it is not possible to just integrate over the entire range
of the inner variable because the integral of the last term on the right 
hand side is divergent:
\beq
  \int_0^\infty 
   \left( \lim_{Z \rightarrow 0} \frac{\partial \Fb_n}{\partial Z} \right)
   dz = 
   \left ( \lim_{Z \rightarrow 0} \frac{\partial \Fb_n}{\partial Z} \right)
   \int_0^\infty dz.
\eeq
However, because the boundary layer has an $O(\eps)$ width and all of
the integrands are $O(1)$, we expect that all of the integrals should be
$O(\eps)$ quantities.

The problem with the intuitive approach is that it makes the mistake of
equating the asymptotic limit $\eps \rightarrow 0$ with the spatial
limit $z \rightarrow \infty$.  Realizing this subtle distinction
(which is safe to neglect for many asymptotic analyses), we can reformulate
the integration over the boundary layer as the limit of a sequence of
integrals over finite intervals, which tends to the
entire half-space $[0,\infty)$ as $\eps \rightarrow 0$.
In choosing the domain of integration, we want to be sure to capture
the entire boundary layer so that the notion of the total surface
excess concentration is physically meaningful.
In addition, we want the region of integration at the macroscopic length 
scale to go to zero as $\eps \rightarrow 0$ so that we are truly integrating 
over only the boundary layer.  We can simultaneously achieve both of these 
goals by taking the region of integration at the macroscopic length
scale (\ie where $Z$ is the variable in the normal direction)
to be $[0,\alpha]$ where $\alpha = O(\eps^{p})$ with $0 < p < 1$. 
Since the width of the boundary layer is $O(\eps)$, this
choice of $\alpha$ ensures that the integration region completely
covers the boundary layer but tends to $0$ in the asymptotic limit.

Even with this choice of integration region, we still must exercise
care to make sure that all integrands are of excess quantities so that 
integrations over the inner coordinate are convergent.  This restriction
suggests that we rewrite (\ref{eq:gamma_evolution_equation}) as
\bea
  \frac{\partial \gamma}{\partial t} = 
  -\nabla_s \cdot \left ( \Fh_s - \F^*_s \right )
  -\frac{1}{\eps^2} \frac{\partial \Fh_n}{\partial z} 
  +\lim_{Z \rightarrow 0} \frac{\partial \Fb_n}{\partial Z}.
  \label{eq:gamma_evolution_eqn_excess_form}
\eea
Integrating this equation over the boundary layer (at the macroscopic 
length scale), we obtain
\bea
  \int_0^\alpha \frac{\partial \gamma}{\partial t} dZ &=&
  - \int_0^\alpha \nabla_s \cdot \left ( \Fh_s - \F^*_s \right ) dZ 
  - \frac{1}{\eps^2} \int_0^\alpha \frac{\partial \Fh_n}{\partial z} dZ
  + \int_0^\alpha 
    \left( \lim_{Z \rightarrow 0} \frac{\partial \Fb_n}{\partial Z} \right)
    dZ.
\eea
Changing from the outer to the inner coordinate for all of the integrals 
except the last term yields
\bea
   \eps \int_0^{\alpha/\eps} \frac{\partial \gamma}{\partial t} dz &=&
   -\eps \int_0^{\alpha/\eps} \nabla_s \cdot \left ( \Fh_s - \F^*_s \right ) dz
   -\frac{1}{\eps} \int_0^{\alpha/\eps} 
       \frac{\partial \Fh_n}{\partial z} dz
  + \int_0^\alpha 
    \left( \lim_{Z \rightarrow 0} \frac{\partial \Fb_n}{\partial Z} \right)
    dZ
  \nonumber \\
  &=&
  -\eps \int_0^{\alpha/\eps} \nabla_s \cdot \left ( \Fh_s - \F^*_s \right ) dz
  -\frac{1}{\eps} \Fh_n(x,y,\alpha/\eps) 
  +\alpha
    \left( \lim_{Z \rightarrow 0} \frac{\partial \Fb_n}{\partial Z} \right)
  \label{eq:integral_inner_coordinate}
\eea
Note that to move from the first to the second line in the above equations, 
we have explicitly integrated the normal derivative and applied the
no flux boundary condition, $\frac{\partial \Fh}{\partial n}(z=0) = 0$.
Expanding the last term in (\ref{eq:integral_inner_coordinate}) 
using a Taylor series, we find that 
\bea
 \eps \int_0^{\alpha/\eps} \frac{\partial \gamma}{\partial t} dz &=&
  - \eps \int_0^{\alpha/\eps} \nabla_s \cdot \left ( \Fh_s - \F^*_s \right ) dz
  - \frac{1}{\eps} \Fh_n(x,y,\alpha/\eps)
  \nonumber \\ 
  & & +~\Fb_n(X,Y,\alpha) - \Fb_n(X,Y,0) + O(\alpha^2).
  \label{eq:taylor_expand_bulk_flux}
\eea
Finally, recalling that $\Fh_n (x,y,\alpha/\eps) = \eps \Fb_n (X,Y,\alpha)$, 
the above equation can be simplified to
\bea
 \eps \int_0^{\alpha/\eps} \frac{\partial \gamma}{\partial t} dz &=&
  - \eps \int_0^{\alpha/\eps} \nabla_s \cdot \left ( \Fh_s - \F^*_s \right ) dz
  - \Fb_n(X,Y,0) + O(\alpha^2).
\eea
By choosing $1/2 < p$ in the definition of $\alpha$, we find 
that the $O(\alpha^2)$ term becomes negligible compared to the remaining
terms in the $\eps \rightarrow 0$ limit so that the leading order
asymptotic equation describing surface concentration evolution satisfies:
\bea
 \eps \int_0^\infty \frac{\partial \gamma}{\partial t} dz &=&
  - \eps \int_0^{\infty} \nabla_s \cdot \left ( \Fh_s - \F^*_s \right ) dz
  - \Fb_n(X,Y,0) 
  \label{eq:Gamma_definition}
\eea
Thus, by substituting the definition for the surface excess
concentration 
\beq
  \Gamma \equiv \eps \int_0^\infty \gamma dz,
  \label{eq:surf_excess_conc_formula}
\eeq
we can make the identifications 
\beq
\J_s \equiv \eps \int_0^\infty \left( \Fh_s - \F^*_s \right) dz 
\label{eq:J_s_formula}
\eeq
and
\beq
J_n \equiv \Fb(X,Y,0) \cdot \hat{n} = -\F^*_n
\label{eq:J_n_formula}
\eeq
to arrive at the 
surface conservation law (\ref{eq:surface_conservation_law}).
The sign difference between $J_n$ and $\Fb_n$ is merely a byproduct of
the choice of orientation for the local coordinate system in our analysis.
As mentioned earlier, the sign convention for the normal flux is that
$J_n$ be positive when the direction of the flux is into the boundary 
layer.

It is worth mentioning that the presence of the $\eps$ in the time 
dependent term and the surface flux term in (\ref{eq:Gamma_definition}) 
indicates that the relative importance of these terms relative to bulk 
transport (\ie the normal flux term) may depend on the choice of time-scales 
and the magnitude of surface transport.   This observation is elaborated 
upon in the examples discussed below.

\section{Formulation in Terms of Chemical Potentials 
\label{sec:formulation_chem_potential}}
In general, it best to express driving forces for fluxes in terms of 
gradients of chemical potentials~\cite{newman_book,probstein_book}:
\beq
  \F_i = -c_i \left( \sum_j L_{ij} \nabla \mu_j \right) + u c_i,
  \label{eq:general_flux}
\eeq
where the $L_{ij}$ are mobility coefficients\endnote{In dimensional
form, mobility coefficients are related to diffusion coefficients via the 
Einstein relation: $D_{ij} = kT L_{ij}$.  With our choice of energy scale, 
the dimensionless Einstein relation is $D_{ij} = L_{ij}$.} that relate the 
drift velocity of species $i$ to the gradient of the chemical potential of 
species $j$ and $u$ is the background velocity that contributes to advective 
transport.  Note that in order to satisfy (\ref{eq:inner_outer_flux_relation}),
we require that the normal component of the background velocity, $u_n$, at the 
interface must vanish.

Substituting this expression into (\ref{eq:J_s_formula}) and 
(\ref{eq:J_n_formula}) and rearranging a bit, we obtain
\bea
  \J^i_s &=& 
     - \Gamma_i \left( \sum_j L_{ij} \nabla_s \mu^*_j \right) 
     + \Gamma_i u^*_s + \epsilon\int_0^\infty \ch_i (\hat{u}_s - u^*_s) dz
     \nonumber \\
   & & -\ \epsilon\int_0^\infty \ch_i \nabla_s 
       \left( \sum_j L_{ij} \left(\hat{\mu}_j - \mu^*_j \right) \right) dz
    \label{eq:general_fluxes_s_with_int} \\
  J^i_n &=& -c^*_i 
     \left( \sum_j L_{ij} \frac{\partial \mu^*_j}{\partial n} \right) 
  \label{eq:general_fluxes_n},
\eea 
where we have imposed $\hat{u}_n = 0$, 
$u_s$ is the tangential component of the background velocity, 
and the superscript $i$ indicates on the $\J_s$ and $J_n$ indicates
that these fluxes are for the $i$-th species.

The expression for $\J^i_s$ can be simplified because the last integral 
term is equal to zero.  As (\ref{eq:conservation_law_inner}) shows, a 
rescaling of the conservation law associated with (\ref{eq:general_flux}) 
to the inner coordinate yields the leading order equation:
\beq
  \frac{\partial}{\partial z} 
    \left[ c_i \left( 
       \sum_j L_{ij} \frac{\partial \hat{\mu}_j}{\partial z} \right)
    \right] = 0.
  \label{eq:quasi_equilibrium}
\eeq
Integrating this equation and using asymptotic matching for the boundary
conditions as $z \rightarrow \infty$, we find that at leading order,
\beq 
  \sum_j L_{ij} \hat{\mu}_j = \sum_j L_{ij} \mu^*_j
  \label{eq:quasi_equilibrium_bdry_layer_eqn}
\eeq
in the inner layer.  That is, special linear combinations of the chemical 
potentials involving the mobility coefficients are constant in the normal 
direction and slowly varying in the tangential direction.  
Equation (\ref{eq:quasi_equilibrium}) has the following physical 
interpretation:
in the sharp interface limit (\ie $\epsilon \rightarrow 0$), the interfacial 
layer is \emph{always} in quasi-equilibrium at the bulk diffusion time scale, 
$\tau = L^2/D$.  It is important to note that at faster time scales, 
which may be appropriate for systems with external forcing, this result may
break down because the time derivative term could balance the normal 
flux term when (\ref{eq:conservation_law_inner}) is rescaled to the faster 
time scale.

Using this observation, $\J^i_s$ becomes
\beq
  \J^i_s =
     - \Gamma_i \left( \sum_j L_{ij} \nabla_s \mu^*_j \right) 
     + \Gamma_i u^*_s + \epsilon\int_0^\infty \ch_i (\hat{u}_s - u^*_s) dz
    \label{eq:general_fluxes_s}.
\eeq
The surface conservation law for transport follows directly from these 
results:
\bea
  \frac{\partial \Gamma_i}{\partial t} &=&
     \nabla_s \cdot \left[ 
        \Gamma_i \left( \sum_j L_{ij} \nabla_s \mu^*_j \right) 
     + \Gamma_i u^*_s 
     + \epsilon\int_0^\infty \ch_i (\hat{u}_s - u^*_s) dz \right ]
    \nonumber \\
    &-& c^*_i 
     \left( \sum_j L_{ij} \frac{\partial \mu^*_j}{\partial n} \right) 
  \label{eq:surf_cons_law_general_transport}
\eea
It is worth mentioning that the surface transport term (the term involving
the surface divergence) does not always contribute to the leading order 
surface conservation law.  Whether the surface transport term must be 
retained at leading order depends on the magnitudes of $\Gamma_i$ (which 
depends implicitly on bulk field variables) and the tangential component of 
bulk chemical potential gradients.  Interestingly, when the surface 
transport term is significant, the surface conservation law depends 
explicitly on the small parameter~$\eps$ (through $\Gamma_i$).

\subsection{Theoretical Derivation of Interfacial Structure}
While surface conservation laws may be derived by substituting 
\emph{arbitrary} models for the structure of the boundary layer into 
(\ref{eq:surf_excess_conc_formula}), (\ref{eq:J_s_formula}), 
(\ref{eq:J_n_formula}), the quasi-equilibrium nature of the interfacial 
layer allows us to theoretically \emph{derive} the boundary layer structure 
from the chemical potential using (\ref{eq:quasi_equilibrium_bdry_layer_eqn}).  
Starting with a free energy for the system that incorporates the physical 
effects we wish to include, chemical potentials may be easily computed as 
the functional derivatives of the free energy with respect to the 
concentrations of the individual species\endnote{This expression for the 
chemical potential implicitly assumes that the system is \emph{locally} in 
thermal equilibrium.  Given this assumption, it is easy relate 
(\ref{eq:def_chemical_potential}) to the definition from equilibrium 
thermodynamics by recognizing that the functional derivative of the total 
free energy with respect to concentration is merely the partial derivative 
of the free energy density with respect to concentration, which yields the 
local chemical potential of the system at each point in space.}:
\beq
  \mu_i = \delta F/\delta c_i,
  \label{eq:def_chemical_potential}
\eeq
where $F$ is the total free energy of the system.
Thus, formulating the transport equations in terms of chemical potentials 
yields a systematic method for deriving boundary layer structure and surface 
conservation laws directly from fundamental physical principles without 
resorting to \emph{ad-hoc} models for the interfacial layer.  We shall
demonstrate this general approach for several example problems in the next 
few sections.

\section{Applications to Neutral Solutes}
In this section, we present surface conservation laws for neutral 
solutions.  Because we have formulated surface conservation laws in 
terms of chemical potentials, derivation of the integrated fluxes for 
these systems is a straightforward application of the formulae in 
Section~\ref{sec:formulation_chem_potential}.  For simplicity, we shall 
assume that fluid flow is absent.

\subsection{Dilute Solutions}
For dilute solutions of a single neutral species, we can write the free 
energy, $F$, for the system as
\beq
 F = U-TS = \int c \Phi 
   + c \left( \log c - 1 \right) \ dx,
 \label{eq:neutral_solution_dilute_free_energy}
\eeq
where $U$, $T$, and $S$ are the internal energy, the absolute temperature 
(scaled to $1$ in dimensionless units), and entropy of the system, 
respectively, $c$ is the concentration of solute, and $\Phi$ is the energy 
of interaction between solute particles and the surface (per unit concentration 
of solute particles). 
Note that the free energy density per unit concentration of solute in 
(\ref{eq:neutral_solution_dilute_free_energy}) is determined up to an 
arbitrary additive constant, which we take to be $-1$ ($-kT$ in dimensional
form) to simplify the expression for the chemical potential (derived below).
The interaction energy, $\Phi(z)$, between solute particles and the surface 
extends over a distance $\epsilon$ (typically a few molecular diameters) and 
can account for hydrophobic interactions, polarization effects, etc.  
$\Phi(z)$ essentially gives the affinity of the solute particles for the 
surface.  Note that in the dilute solution approximation, solute particles 
do not directly interact with each other or with solvent particles.

Computing the functional derivative of 
(\ref{eq:neutral_solution_dilute_free_energy}), we obtain an expression
for the chemical potential 
\beq
  \mu =  \log c + \Phi.
  \label{eq:neutral_solution_dilute_chemical_potential}
\eeq
As we showed in Section~\ref{sec:formulation_chem_potential}, in the
interfacial layer, $\hat{\mu}$ is equal to $\mu^*$ in the normal direction, 
which leads to a Boltzmann distribution for the concentration:
\beq
  \ch(z) = \exp \left( \mu^*-\Phi(z) \right).
  \label{eq:neutral_solution_dilute_conc_profile}
\eeq
We can rewrite this expression in terms of the bulk concentration just 
outside of the interfacial layer by observing that
$c^* = e^{\mu^*}$ from
asymptotic matching and the fact that the solute surface interaction
decays far away from the surface.  Thus, we find that
\beq
  \ch(z) = c^* e^{-\Phi(z)}.
  \label{eq:neutral_solution_dilute_conc_profile_alt}
\eeq
Note that this is exactly the boundary layer profile obtained by 
Anderson~\etal~using the \emph{assumption} that the interfacial layer is 
in equilibrium~\cite{anderson1982}.

The surface excess concentration of solute, $\Gamma$, is the integral of the
excess concentration over the boundary layer:
\beq
  \Gamma = \epsilon \int_0^\infty \left( \ch(z) - c^* \right) dz 
  = \epsilon \int_0^\infty c^* 
      \left (e^{-\Phi(z)} - 1 \right) dz
  \label{eq:neutral_solution_dilute_gamma}.
\eeq
After the integral in this expression has been explicitly evaluated, the 
surface conservation law for dilute solutions of neutral solutes is
derived by substituting the surface excess concentration
(\ref{eq:neutral_solution_dilute_gamma}) and the chemical potential
(\ref{eq:neutral_solution_dilute_chemical_potential})
into (\ref{eq:surf_cons_law_general_transport}), which leads to 
the following transport equation for the surface excess concentration:
\beq
  \frac{\partial \Gamma}{\partial t} = 
    L \left[
      \nabla_s \cdot \left( \Gamma \ \nabla_s \log c^* \right)
    - \frac{\partial c^*}{\partial n} \right],
  \label{eq:neutral_solution_dilute_scl}
\eeq
where $L$ is the nondimensional mobility coefficient for solute particles.
It is interesting to observe that the driving force for surface transport 
and the source term for transport between the bulk and interfacial regions 
depend only on the concentration in the bulk.  This result is generally
true for transport problems where fluid flow is negligible.

\subsection{Concentrated Solutions}
For concentrated solutions of a single neutral species, the size of the
individual solute particles cannot be neglected and steric effects must be
accounted for.  Concentrated solution theory provides a model which 
qualitatively accounts for these effects.  Bulk solutions are considered 
concentrated when the dimensionless parameter $\nu = a^3 c_{max}$, which 
represents the bulk volume fraction of solute, is not small.  In this case, 
we must use concentrated solution theory to describe bulk transport.  However, 
even when $\nu \ll 1$, concentrated solution theory may be necessary if the 
surface interaction energy, $\Phi$, is strong enough to condense particles 
near the surface.  As for dilute solutions, we begin by writing the free 
energy for the system using ideal solution theory~\cite{lupis_book}: 
\beq
  F = U-TS = \int c \Phi 
    + \left [ c \log c + (1-c) \log \left(1 - c \right) \right] \ dx,
  \label{eq:neutral_solution_conc_free_energy}
\eeq
where the last term in $F$ accounts for then entropy of the solvent 
(\ie steric effects) and we have assumed that solute and solvent 
molecules have the same size.  Taking the functional derivative of this
expression, we find that the chemical potential for concentrated solutions
is given by 
\beq
  \mu = \log \left ( \frac{c}{1 - c} \right) + \Phi.
  \label{eq:neutral_solution_conc_chemical_potential}
\eeq
Note that unlike dilute solutions, there is a maximum concentration for
the solute that arises from the sharp increase in the chemical potential
at high solute concentrations (low solvent concentrations) caused
by the solvent's entropic contribution to the free energy.

As before, we can derive the concentration profile of solute atoms by
using the fact that, within the interfacial layer, the chemical potential 
is constant in the direction normal to the surface: 
\beq
  \ch(z) = 
    \exp \left(\mu^* - \Phi(z) \right)
    \left[ 
      1 + \exp \left( \mu^* - \Phi(z) \right)
    \right]^{-1}.
  \label{eq:neutral_solution_conc_conc_profile}
\eeq
For concentrated solutions, the simple Boltzmann distribution for the 
quasi-equilibrium concentration profile in the normal direction is replaced
by a Fermi-Dirac-like distribution with a maximum concentration of $1$
($c_{max} = 1/a^3$ in dimensional form).
Note that expression of $\ch(z)$ in terms of the bulk concentration, 
$c^*$, outside of the interfacial layer does not lead to a significant 
simplification of (\ref{eq:neutral_solution_conc_conc_profile}).  
The surface excess concentration is found by integrating 
(\ref{eq:neutral_solution_conc_conc_profile}) over the interfacial layer.

To derive the surface conservation law for concentrated neutral solutions,
we substitute (\ref{eq:neutral_solution_conc_chemical_potential})
and the expression for the surface excess concentration into the general
formula for the surface conservation law in terms of chemical potentials
(\ref{eq:surf_cons_law_general_transport}).

\section{Application to Electrolytes}
In electrochemical systems, the solute particles are electrically charged, 
so transport processes are affected by electric fields (both self-generated
and externally applied).  When modeling electrochemical systems, it is 
common to model transport in the bulk separately from the response of the
double layer at surfaces.  Surface conservation laws justify and generalize
the usual phenomenological approaches for coupling bulk and double layer 
dynamics.  In this section, we discuss surface conservation laws for dilute
and concentrated electrolyte solutions.  As we shall see, the key difference
between these two cases is the microscopic model that must be used
to describe the electrical double layer.  

For electrolytes, transport is driven by gradients in the electrochemical
potential, so the nondimensionalized conservation laws for ionic species are 
given by
\beq
  \frac{\partial c_i}{\partial t} = 
    \nabla \cdot \left( c_i \nabla \mu_i + c_i v \right) 
  \label{eq:NP_eqns_mu},
\eeq
where $c_i$ are the concentration of ion $i$, $\mu_i$ is the electrochemical 
potential for species $i$, $v$ is the fluid velocity,
and we have assumed that all of the ions have the same 
diffusivity~\cite{newman_book}.
Note that the flux in this equation is just a special case of 
(\ref{eq:general_flux}) where $L_{ij}$ is an identity matrix\endnote{In
dimensional form, $L_{ij}$ would be a scalar multiple of the identity matrix.}.
While fluid flow plays an important role in many systems of current 
interest (\eg electrokinetic microfluidic pumps and 
mixers~\cite{ramos1999,ajdari2000,iceo2004a,iceo2004b}), we 
shall neglect it to keep the following discussion simple.  In this 
situation, the surface conservation law 
(\ref{eq:surf_cons_law_general_transport}) becomes
\beq
  \frac{\partial \Gamma_i}{\partial t} = 
  \nabla_s \cdot \left( \Gamma_i \nabla_s \mu^*_i \right) 
    - c^* \frac{\partial \mu^*_i}{\partial n}.
  \label{eq:surf_cons_law_ion_transport}
\eeq

Before considering ion transport in various special cases, we mention that
for electrochemical systems, the small parameter $\epsilon$ is the ratio
of the Debye screening length, $\lambda_D$, to the characteristic
system size.  In this article, the Debye length will be defined by
\beq
  \lambda_D = \sqrt{\frac{\epsilon_s k T}{2 e^2 C_o}},
  \label{eq:debye_length}
\eeq
where $\epsilon_s$ is the dielectric constant for the electrolyte
and $C_o$ is the average concentration of neutral salt.  Alternative choices 
for the Debye length typically differ from this definition by an $O(1)$ 
multiplicative constant.

\subsection{Dilute Electrolytes}
The free energy for general dilute electrolyte solutions is similar to the 
free energy for dilute neutral solutions: 
\beq
  F = U - TS = \int \left (
      \sum_i  c_i \left ( \log c_i - 1 \right )
    + \sum_i  z_i c_i \phi 
    - \epsilon^2 |\nabla \phi|^2
  \right ) dx.
  \label{eq:dilute_electrolytes_free_energy}
\eeq
In this expression, the first term in the integrand is the entropic 
contribution from the solute particles, the second term accounts for the 
interaction energy between the charged particles and the electric potential, 
and the last term is the energy of the electric field.  Note that 
$\epsilon$ is defined to be consistent with the definition in 
(\ref{eq:debye_length}).

Taking the functional derivative of the free energy with respect to $c_i$, 
we find that the electrochemical potential of the $i$-th ionic species takes 
the simple form 
\beq
  \mu_i = \log c_i + z_i \phi.
  \label{eq:dilute_electrolytes_electrochem_pot}
\eeq
Using (\ref{eq:dilute_electrolytes_electrochem_pot}) for the electrochemical 
potential, (\ref{eq:NP_eqns_mu}) reduces to the commonly used Nernst-Planck 
equations~\cite{bard_book, rubinstein_book}:
\beq
  \frac{\partial c_i}{\partial t} =
    \nabla \cdot \left( \nabla c_i + z_i c_i \nabla \phi \right) 
  \label{eq:NP_eqns}
\eeq
Within the double layer, the electrochemical potential is constant in the 
normal direction, so the concentration profile possesses a Boltzmann 
distribution:
\beq
  \ch_i(z) = c^*_i e^{- z_i \psih(z)},
  \label{eq:dilute_electrolytes_conc_profile}
\eeq
where $\psih(z) \equiv \phih(z) - \phi^*$. 

For the case of a symmetric, binary electrolyte, 
(\ref{eq:dilute_electrolytes_conc_profile}) leads to the commonly used 
Gouy-Chapman-Stern (GCS) model for the double 
layer~\cite{bard_book, hunter_book, newman_book, chu2005}.
It is well known that the GCS model can be theoretically justified by
noting that the double layer is always in quasi-equilibrium in the 
sharp interface limit~\cite{bazant2005}.  However, the quasi-equilibrium 
structure of the double layer is not at all surprising in light of our 
discussion in Section~\ref{sec:formulation_chem_potential} -- it follows 
directly from the fact that the electrical double layer is a sharp 
interfacial layer. 

For symmetric, binary electrolytes, we can derive explicit expressions for 
the excess surface concentrations.  In this case, the excess concentration 
of each of the two ionic species is given by 
\beq
  \gamma_\pm = \ch_\pm - c^*_\pm = c^* \left ( e^{\mp z_{_+} \psih} - 1 \right )
  \label{eq:dlc_gamma}
\eeq
where $c^* = (c^*_+ + c^*_-)/2$ is the average concentration in the bulk
and $z_{_+}$ is the charge number for the positive ion.
Following~\cite{chu2006}, this quantity is straightforward to integrate by 
changing the variable of integration in (\ref{eq:surf_excess_conc_formula}) 
to $\psih$ and using the fact that 
\beq
  \frac{\partial \psih}{\partial z} = 
    -2 \sqrt{c^*} \sinh \left( \frac{z_{_+} \psih}{2} \right)
  \label{eq:dlc_dpsi_dz}
\eeq
for the GCS model.
Carrying out the integration, we find that the excess surface concentration 
of species $i$ is given by
\beq
  \Gamma_\pm = \frac{2 \eps \sqrt{c^*}}{z_{_+}} 
    \left( e^{\mp z_{_+} \zeta/2} - 1 \right)
  \label{eq:dlc_Gamma}
\eeq
where $\zeta = \phih(0) - \phi^*$ is the 
zeta-potential~\cite{hunter_book, bard_book, newman_book} 
across the diffuse part of the double layer.
Using this expression for $\Gamma_\pm$ in 
(\ref{eq:surf_cons_law_ion_transport}), the surface conservation law for 
symmetric, binary electrolytes is fully specified in terms of bulk field 
variables and boundary conditions, and the structure of the boundary layer 
has been completely integrated out.

Because the charge density and neutral salt concentration are both important
components of the response of an electrochemical system, it is interesting to 
derive surface conservation laws for 
these quantities~\cite{bazant2004,chu2006}.
Toward this end, we define $\eps q$ and $\eps w$ to be the surface charge 
density and surface excess neutral salt concentration, 
respectively\endnote{The factor of $1/2$ in the definition of $q$ is 
present for mathematical convenience.  The total surface charge density 
is $2\eps q$.}:
\bea
  \eps q &=& 
     \frac{\eps}{2} \int_0^\infty \left( \gamma_+ - \gamma_- \right) dz 
     = \frac{1}{2} \left( \Gamma_+ - \Gamma_- \right)
     = -\frac{2 \eps \sqrt{c^*}}{z_+} \sinh \left( \frac{z_+ \zeta}{2} \right)
  \label{eq:dlc_q_def} \\
  \eps w &=& 
     \frac{\eps}{2} \int_0^\infty \left( \gamma_+ + \gamma_- \right) dz 
     = \frac{1}{2} \left( \Gamma_+ + \Gamma_- \right)
     = \frac{4 \eps \sqrt{c^*}}{z_+} \sinh^2\left( \frac{z_+ \zeta}{4} \right).
  \label{eq:dlc_w_def}
\eea
Using (\ref{eq:dlc_q_def}) and (\ref{eq:dlc_w_def}), we can combine the 
surface conservation laws for individual ions 
(\ref{eq:surf_cons_law_ion_transport}) and write out the electrochemical
potential in terms of $c^*$ and $\phi^*$ to obtain
\bea
  \eps \frac{\partial q}{\partial t} &=&
     \eps \nabla_s \cdot 
     \left(
         q \nabla_s \log c^*
       + z_+ w \nabla_s \phi^*
     \right)
     - z_+ c^* \frac{\partial \phi^*}{\partial n} 
  \label{eq:q_evolution_eqn} \\
  \eps \frac{\partial w}{\partial t} &=& 
     \eps \nabla_s \cdot 
     \left(
         w \nabla_s \log c^*
       + z_+ q \nabla_s \phi^*
     \right)
     - \frac{\partial c^*}{\partial n}.
  \label{eq:w_evolution_eqn} 
\eea
Since we have explicit expressions for $q$ and $w$, (\ref{eq:q_evolution_eqn}) 
and (\ref{eq:w_evolution_eqn}) are completely specified in terms of bulk 
field variables yielding surface conservation laws that can serve as
effective boundary conditions for the bulk equations written in terms of 
these field variables.

\subsection{Concentrated Electrolytes}
Following Iglic and Kralj-Iglic~\cite{iglic1994,kralj-iglic1996}
and Borukhov \etal~\cite{borukhov1997,borukhov2000,borukhov2004},
we can write a free energy that accounts for steric interactions:
\bea
  F &=& U - TS \nonumber \\
    &=& \int \left (
      \sum_i  c_i \log c_i 
    + \left ( 1 - \sum_i  c_i \right ) \log \left( 1 - \sum_i  c_i \right) 
    + \sum_i  z_i c_i \phi 
    - \epsilon^2 |\nabla \phi|^2
  \right ) dx.
  \label{eq:conc_electrolytes_free_energy}
\eea
As for concentrated neutral solutions, steric effects are included by 
adding a contribution from the solvent entropy (second term in integrand)
and we have assumed that the molecules of all species present in the 
system (including solvent molecules) are of the same size.  
It is important to mention that using (\ref{eq:conc_electrolytes_free_energy})
as the free energy is only one way of including steric effects.  There
are more sophisticated (and accurate) theories involving statistical density
functional 
theory~\cite{gillespie2002,gillespie2003,gillespie2005,henderson2005}.
Unfortunately, these theories are generally more cumbersome to analyze
and often require advanced numerical methods in order to gain physical
insight~\cite{kilic2007a}. 

Next, we derive electrochemical potentials for the $i$-th ionic species 
following Kilic \etal~\cite{kilic2007b} by taking the functional 
derivative of the free energy with respect to the $i$-th concentration:
\beq
  \mu_i = \log \left( \frac{c_i}{1-\sum_j c_j} \right) + z_i \phi.
  \label{eq:conc_electrolytes_electrochem_pot}
\eeq
This expression for the electrochemical potential leads to a modified
Nernst-Planck equation for ion transport~\cite{kilic2007b}
\beq
  \frac{\partial c_i}{\partial t} =
    \nabla \cdot \left( 
      \nabla c_i 
      + z_i c_i \nabla \phi 
      + \left[\frac{c_i}{1-\sum_j c_j}\right] \sum_j \nabla c_j
    \right) 
  \label{eq:modified_NP_eqns}.
\eeq
Notice that, unlike dilute electrolytes, the transport of different 
ionic species are directly coupled via the last term in the flux.

Following our previous procedure, we can derive the concentration profile of 
each ionic species within the double layer by using the fact that each of the
$\mu_i$ must be constant in the normal direction.  This observation leads
to a linear system of equations for the ionic concentrations, which is 
easily solved to yield
\beq 
  \ch_i(z) = \frac{\exp \left (\mu^*_i - z_i \phih(z) \right)}
               {1 + \sum_j \exp \left( \mu^*_j - z_j \phih(z) \right) }
  \label{eq:conc_electrolytes_conc_profile}
\eeq
As for concentrated neutral solutions, the concentration profiles 
have a Fermi-Dirac-like form. 

A little further progress can be made for the special case of symmetric, 
binary electrolytes.  The excess concentrations of the ionic species in 
this case are given by
\beq
  \gamma_\pm = \ch_\pm - c^*_\pm 
    = c^* \left[ \frac{ e^{\mp z_{_+} \psih}}
     {1+2 \nu \sinh^2\left( z_{_+} \psih/2 \right)} - 1 \right],
  \label{eq:conc_electrolytes_gamma}
\eeq
where $c^*$, $\psih$ and $\nu$ are as defined in our discussion of dilute
electrolytes and concentrated neutral solutions.
Unfortunately, integration of these excess concentrations for the 
concentrated electrolyte model does not yield a particularly simple 
expression for the surface excess concentration~\cite{kilic2007a}:
\bea
  \Gamma_{\pm} &=& 
    \mp \ \sgn{\zeta} z_{_+} \epsilon
    \sqrt{\frac{2}{\nu} \log \left( 1 + 2\nu \sinh^2(z_{_+} \zeta/2) \right) }
  \nonumber \\
  &+& (1-\nu) \epsilon \int_0^{z_{_+}\zeta}
      \frac{\cosh u - 1}{1+2 \nu \sinh^2 u}
      \left[ \frac{2}{\nu} 
        \log \left( 1 + 2\nu \sinh^2 u \right) 
      \right]^{-1/2}
      du.
\eea
Combining this expression for the surface excess concentration and the 
electrochemical potential for concentrated electrolytes 
(\ref{eq:conc_electrolytes_electrochem_pot}) yields the surface conservation
law for concentrated, symmetric binary electrolytes.

\subsection{Dominant Terms in Electrochemical Surface Conservation Laws}
In general, the relative importance of the terms in a set of surface
conservation laws (written in any form) may depend on the choice of 
time scales and the magnitude of surface transport.  
In the context of electrochemical transport, we find that the boundary
conditions applied in many theoretical studies are merely the leading order
form (in different asymptotic and physical regimes) of the surface 
conservation laws derived in the previous section.
For instance, in induced charge electro-osmosis problems in dilute solutions
at weak applied electric fields~\cite{iceo2004a,iceo2004b}, $\eps q$ and 
$\eps w$ remains $O(\eps)$ quantities.  
Thus, compared to the normal flux term, the surface flux term is negligible 
and the time-dependent term is only important at short times, $t = O(\eps)$. 
In this situation, the double layer charging equation 
(\ref{eq:q_evolution_eqn}) becomes \cite{iceo2004b, iceo2004a}:
\beq
  \frac{\partial q}{\partial \tilde{t}} = 
  \sigma \frac{\partial \phi}{\partial n},
    \label{eq:iceo_charging_equation}
\eeq
where $\sigma = c^*$ is the bulk conductivity of the solution and 
time has been rescaled using $\tilde{t} = t / \eps$ so that the dynamics 
are on the RC time scale \cite{iceo2004b,bazant2004,iceo2004a}.  
At $t = O(1)$ time scales, only the normal flux remains an $O(1)$ quantity, 
so we are left with a ``insulator'' boundary condition for the electric
potential:
\beq
  \sigma \frac{\partial \phi}{\partial n} = 0.
    \label{eq:iceo_charging_equation_long_times}
\eeq

At higher applied fields or for highly charged
particles~\cite{dukhin1969,hinch1983,hinch1984,shilov1970}, the dilute
solution electrolyte model leads to surface excess concentrations as large 
as $O(1/\eps)$ for some of the ionic species in the double layer.  As a 
result, surface currents becomes important and evolution of ionic 
concentrations within the double layer occurs on an $O(1)$ time scale.  In 
this situation, no terms in (\ref{eq:surf_cons_law_ion_transport}) are 
negligible, so we must retain all terms in (\ref{eq:q_evolution_eqn}) and
(\ref{eq:w_evolution_eqn}).

\section{Summary}
In this article, we have presented a general formulation and derivation of 
surface conservation laws at microscopically diffuse interfaces.  Because
surface conservation laws form the crucial connection between bulk
and interfacial dynamics, it is important to know that they are 
theoretically well-founded.  Our work fills this apparent void in the 
literature and provides a solid theoretical foundation based on techniques
from asymptotic analysis.  Our analysis has also led to explicit formulae for 
the surface and normal fluxes involved in surface conservation laws.  
In addition to formulae for arbitrary interfacial models, we have 
presented a formulation of surface conservation laws for the important class 
of interfacial models derived using the principles of non-equilibrium
thermodynamics.  This formulation provides a method for developing 
interfacial models in a systematic and physically sound manner.
Finally, we have demonstrated the derivation of surface conservation laws
in the specific contexts of diffusive and electrochemical transport.  We 
emphasize, however, that surface conservation laws are very general and 
apply to a wide-range transport processes.   The basic approach is to first 
specify (or derive) a model for the interfacial layer and then to compute 
the surface excess concentrations and surface flux using the general formulae 
presented in this article.

\section*{Acknowledgments}
This work was supported in part by the Department of Energy Computational
Science Graduate Fellowship Program of the Office of Science and
National Nuclear Security Administration in the Department of Energy
under contract DE-FG02-97ER25308 (KTC) and by the MRSEC Program of the
National Science Foundation under award number DMR 02-13282 (MZB).
The authors thank A. Ajdari for helpful discussions on the use of chemical
potentials.

\begingroup
\theendnotes
\endgroup


\providecommand{\refin}[1]{\\ \textbf{Referenced in:} #1}

\end{document}